\documentclass{article}
\usepackage{epsfig}
\usepackage{epsf}
\usepackage{float}
\usepackage{graphicx}
\usepackage{amssymb}
\usepackage{verbatim}
\usepackage{amsmath}
\usepackage{amsfonts}
\usepackage{mycommand}
\usepackage{jrnl}
\usepackage{fancybox}
\usepackage{fancyheadings}
\usepackage[dvipsnames]{color}
\usepackage{pstricks}
\usepackage{hyperref}

\begin{document}

\markboth{H. Barman and N. S. Vidhyadhiraja}
{Transport and spectra in the half-filled Hubbard model}

\title
{\small \bf TRANSPORT AND SPECTRA IN THE
HALF-FILLED HUBBARD MODEL:
A DYNAMICAL MEAN FIELD STUDY}

\author{\small H.\ BARMAN and N.\ S.\ VIDHYADHIRAJA\\
Theoretical Sciences Unit, Jawaharlal Nehru Centre For
Advanced Scientific Research\\
Jakkur, Bangalore 560064, India\\
hbar@jncasr.ac.in}

\maketitle

\begin{abstract}
We study the issues of scaling and universality in spectral and transport properties 
of the infinite dimensional particle--hole symmetric (half--filled) Hubbard model within dynamical
mean field theory. One of the simplest and extensively used impurity solvers, namely 
the iterated perturbation theory approach is reformulated to avoid problems such as
analytic continuation of Matsubara frequency quantities or calculating multi-dimensional
integrals, while taking full account
of the very sharp structures in the Green's functions that arise close to the Mott 
transitions and in the Mott insulator regime. We demonstrate
its viability for the half-filled Hubbard model.
Previous known results are reproduced within the present approach.
The universal behavior of the spectral functions in the Fermi liquid regime
is emphasized, and adiabatic continuity to the non-interacting limit is demonstrated.
The dc resistivity in the metallic regime is known to be a  
non-monotonic function of temperature with a 
`coherence peak'. This feature is shown to be a universal feature occurring at
a temperature roughly equal to the low energy scale of the system. 
A comparison to pressure dependent dc resistivity experiments on
Selenium doped NiS$_2$ yields qualitatively good agreement.
Resistivity hysteresis across the Mott
transition is shown to be described qualitatively within the present framework. 
A direct comparison of the thermal hysteresis observed in V$_2$O$_3$ with our theoretical
results yields a value of the hopping integral, which we find to be in the range
estimated through first-principle methods. Finally, a systematic study of 
optical conductivity is carried out and the changes in absorption as a result
of varying interaction strength and temperature are identified.

\end{abstract}

%
%---------------------------------------------------------------------------
%---------------------------------------------------------------------------
\section{Introduction}
The Hubbard model (HM)~\cite{hubbard:1,hubbard:2,hubbard:3,gutz:1} has become one of the central 
paradigms in theories
for strongly correlated electron systems. For systems ranging from ultracold 
gases~\cite{ultracold}, transition metal oxides~\cite{mott:1,mott:2}, high temperature 
superconductors~\cite{htc:1,jarrell}, and even 
graphene~\cite{graphene:1,graphene:2}, the HM has been employed in various guises. Thus,
the importance of developing an understanding of this model cannot be overstated. 
The theoretical techniques that have been employed in an attempt to solve
this model are diverse, e.g. Hatree-Fock approximation~\cite{hf:1,hf:2,hf:3}, 
Hubbard's Green's function approximations~\cite{gf:1,gf:2}, 
variational wave function method~\cite{gutz:1,gutz:2}, slave Boson approach~\cite{slaveboson}, etc. 
However, to date the general HM remains unsolved except for the 
one-dimensional case~\cite{lieb}. A lot of progress has been made in recent years in this direction 
through the dynamical mean field theory (DMFT)~\cite{metzner,georges:rvw}, where the lattice model 
is mapped onto an effective single impurity model with a self-consistently 
determined hybridization. Within the DMFT context, a greater interest is being devoted towards   
the inifinite dimensional HM since the mapping becomes exact in the limit 
and simplifies many major issues of many-body approaches. Although the lattice problem reduces to that
of an impurity problem embedded in 
a non-interacting bath, the problem remains 
amenable to a host of techniques such as non-crossing approximation (NCA)~\cite{nca}, 
quantum Monte Carlo(QMC)~\cite{jarrell,georges:qmc,georges:rvw}, 
numerical renormalization group (NRG)~\cite{nrg}, exact diagonalization (ED)~\cite{ed},
iterated perturbation theory (IPT)~\cite{yosida,georges:rvw,ipt:1,ipt:2}, 
local moment approach (LMA)~\cite{lma:1,lma:2}, 
etc, that have been
developed to solve the impurity problem. IPT is a particularly
simple diagrammatic perturbation theory  based approach that has been 
extensively used to study not only the impurity problem, but also a host of other 
lattice based models such as the HM and the periodic Anderson model (PAM)~\cite{hewson}.
It is known to yield qualitatively good results, such as the Mott metal-insulator
transition, heavy fermion behavior etc. However, since it is perturbative
by construction, IPT does not do as well in quantitative terms.
The half-filled HM, has likewise, been a very intensively studied 
model, as a description of the Mott MIT, and IPT has been very successful
in elucidating quite a few conceptual issues~\cite{ipt:1,ipt:2}. 
Nevertheless, we observe a few
important deficiencies in the present implementations of IPT,
specifically for the half-filled HM. IPT, as implemented generally, calculates Matsubara
frequency Green's functions and self-energies. The procedure of analytic
continuation, required to obtain real frequency quantities, is known to
be a numerically difficult problem, which is conventionally treated by 
Pad\'e approximation or the maximum entropy method~\cite{num:recipe,georges:rvw}. The direct real frequency implementation takes two
routes -- in the first,
one has to solve multi-dimensional integrals, which is computationally expensive; 
the other route employs Fast Fourier transforms, which has the disadvantage
that the frequency grid is necessarily uniform, and thus cannot sample the
very sharp structures arising in the Green's functions in the parameter region proximal
to the Mott transition as well as in the Mott insulator regime. This leads to problems
in convergence of the DMFT iterations. In a very recent study, an ad-hoc way of
 dealing with the 
sharp structures was implemented through the addition of a small constant 
imaginary part added to the real frequency~\cite{fathi:jafari}. This
method broadens the poles of the Green's functions and lets one
handle the sharp structures numerically. 
However, it
has the disadvantage that it leads to spurious spectral features, e.g. the insulating phase does not exhibit a clear gap.

In this work, we introduce an alternative real frequency implementation
that overcomes the above limitations, and hence, lets one study the 
properties of interest such as spectral functions etc directly on 
the real axis with an adaptive non-uniform grid. 
In order to solve the DMFT self-consistency equations we start with an initial guess for the self-energy (typically the Hatree-Fock) whereas the host
Green's function is guessed initially in common practice.
We benchmark
our results against known IPT results calculated through analytic continuation
of the Matsubara frequency Green's functions. We find excellent agreement.
Issues of universality and scaling of spectral functions and transport
quantities are emphasized. The paper is organized as follows: The next section
details the model and the reformulation of IPT. The numerical implementation
details are also discussed. Section 3 comprises our theoretical results, 
discussion and comparison to experiments.
We present our conclusions in section 4. 

%------------------------------------------------------------------
%------------------------------------------------------------------
\section{Methods and Formalism}
The single-band HM in standard
notation is given by

\be
\label{eq:hm} \hat
H=-\sum_{<ij>,\sigma}t_{ij}(c_{i\sigma}\y
c_{j\sigma}^{\ph d}+\textrm h.c.)
+\eps_d\sum_{i\sigma}c_{i\sigma}\y
c_{i\sigma}^{\ph d}
+U\sum_i n_{i\ua}n_{i\da}
\ee 
where $t_{ij}$ is the amplitude of hopping
of electrons from site-$i$ to site-$j$ ,
$\eps_d$ is the electron's orbital
energy and $U$
is the on-site Coulomb repulsion. The
operator $c_{i\sig}$ ($c_{i\sig}\y$)
annihilates (creates) an electron of spin
$\sig$ at site $i$. In the limit of
infinite dimensions $(d\ra\infty)$, the hopping
parameter is scaled as $t=t_\ast/\sqrt{z}$
where $z$ is the coordination number of the
lattice. In this work, we choose to work
with the hypercubic lattice (HCL) for which
the non-interacting ($U=0$) density of
states is an unbounded gaussian, i.e.
\be\label{eq:dos} D_0(\eps)=\f{1}{\sqrt\pi 
\ts}\exp(-\eps\sq/{\ts\sq}) \ee
The major simplification within DMFT, which
is exact in $d=\infty$,
is that the
self-energy~\cite{metzner} and vertex function~\cite{khurana}
become momentum independent or spatially local.
The local, retarded Green's function in the paramagnetic 
case is given by
\be
\label{eq:local_gf} G(\om)=\sum_{\bf{k}}
\f{1}{\om^+-\eps_k-\eps_d-\Sig(\om)}\ee
where $\om^+=\om + i\eta$, $\eta\rightarrow 0^+$
and $\Sig(\om)$ is the real-frequency self-energy.
The $\bf{k}$ sum may be transformed to a density of
states integral and thus may be written as
\be\label{eq:gf_ht} G(\om)={\mc H}[\g(\om)]
\ee
where ${\mc H}[z]=\int d\eps\,
D_0(\eps)/(z-\eps)$ is the Hilbert
transform of $z$ with respect to
$D_0(\eps)$ and
$\g(\om)=\omp-\eps_d-\Sig(\om)$.
Since a lattice model can be mapped onto an
impurity model within DMFT with the
self-consistency condition that the
impurity self-energy is the same as the
lattice self-energy, one can find the
host or medium Green's function ${\mc G}(\om)$ for the impurity
through the Dyson equation
\be
\label{eq:dyson} {\mc
G}\inv(\om)=G\inv(\om)+\Sig(\om)
\ee
Solution of the impurity model in terms of
$\mc G(\om)$ would then yield a new $\Sig(\om)$
which when put back in Eq.~\eref{eq:gf_ht} gives
$G(\om)$. Thus, given an impurity solver
technique, one can self-consistently solve
for the Green's functions, and hence 
the self-energy.
Since the vertex function is also momentum
independent within DMFT, the calculation of
conductivity through the current-current
correlation function involves only
single-particle quantities. Thus the optical
conductivity may be calculated by using the
following expression obtained from the Kubo formula~\cite{mahan}.
\ber
\hspace*{-3cm}\sig\omb= \f{\sig_0}{2\pi\sq} \rm{Re}
\nint d\om'\,\f{\nf(\om')-\nf(\om+\om')}{\om} 
\times \non \\ 
\bl \f{G^\ast(\om')-G(\om+\om')}{\g(\om+\om')-\g^\ast(\om')}
-\f{G(\om')-G(\om+\om')}{\g(\om+\om')-\g(\om')}\br
\label{eq:optcond} 
\eer
where $\sig_0=4\pi e\sq t\sq a\sq n/\hbar$ for lattice constant $a$,
electronic charge $e$, and electron density $n$.
As $\om \ra 0$ we obtain the DC conductivity :\\
\be
\label{eq:dccond} \sig_{DC}=\f{\sig_0}{2\pi\sq} {\rm Re}
\int^\infty_{-\infty} d\om\,\left(-\f{d\nf}{d\om}\right)\left[
\f{\pi D(\om)}{\rm{Im}\g(\om)}
+ 2(1-\g(\om)G(\om)\right]
\ee
where $D\omb=-\pim G\omb$ is the spectral
function of the retarded Green's function
$G\omb$ and $\nf\omb=(e^{\beta\om}+1)\inv$ is
the Fermi function ($\beta=1/(k_B T)$).\\

As mentioned earlier, the full solution of
DMFT problem requires an impurity
solver technique. In this paper, we have chosen
IPT for solving the self-consistent
impurity problem. IPT does have a few
drawbacks.  It is based on a simple truncation
of the diagrammatic perturbation theory in $U$ to second
order around the Hartree mean field solution. Thus it
needs to be benchmarked for every new
problem. For the same reason, it is unable
to capture exponentially small energy
scales. Nevertheless, the reason for this
choice is that IPT is technically one of the
simplest approaches that can capture the
Mott transition physics in a qualitatively
correct manner. IPT has been extensively employed by
various groups to study the HM~\cite{ipt:1,ipt:2,georges:rvw} 
and PAM~\cite{raja:epl1,raja:epl2,saso,rozenberg:pam}.
However, the previous implementations suffer from the 
problems mentioned in the introduction section.

We now describe IPT and our implementation of it briefly. The ansatz for
the dynamical part of the self-energy (apart from the static Hartree term)
within IPT is just the
second order term of the perturbative
expansion in $U$ about the Hartree limit,
i.e.
\be
\Sigma_2(\om)=\lim_{i\om\rightarrow\omp}\frac{U\sq}{\beta^2}\sum_{m,p}{\mc
G}(i\om+i\nu_m)
{\mc G}(i\om_p+i\nu_m){\mc G}(i\om_p)
\ee
where $i\om$ and $i\nu$ denote odd and even Matsubara frequencies 
respectively. 
Using the spectral representation $\mc
G(\iom)=\disp\nint d\om'\,\disp\f{D_{\mc G}(\om')}{\iom-\om'}$,
(where $D_{\mc G}(\om) = -{\rm Im}{\mc G}(\om)/\pi$) and 
carrying out the Matsubara sums along with the 
trivial analytic continuation $\iom\ra\omp$, we get
the following expressions for the imaginary
part of the self-energy on the
real frequency axis:
\blgn
D_{\Sig}\omb&=-\pim \Sig\omb \non \\
&=U\sq \nint d\om' D_{\mc G}(-\om')
\bl \nf(\om')\chi(\om+\om')
+ \nf(-\om')\chi(-\om-\om') \br
\label{eq:Dsig} 
\elgn
where 
\be \chi \label{eq:chi}\omb
=\nint d\om' 
D_{\mc G}(\om+\om') D_{\mc G}(\om')
\nf(-\om-\om') \nf(\om') \ee
and we get the real part by using Kramers-Kronig 
transformation:
\be
\label{eq:kkt} \mr{Re} \Sig\omb 
=\frac{{\mc P}}{\pi}\nint d\om'\,
\f{\mr{Im} \Sig (\om')}
{\om'-\om}
\ee
Eq.~\eref{eq:local_gf},
\eref{eq:gf_ht}, and \eref{eq:dyson}
along with Eq.
\eref{eq:Dsig}-\eref{eq:kkt} constitute the necessary
ingredients for the solution of the
half-filled Hubbard model within DMFT. We remark here that
the integrals above are all one-dimensional and hence do not
present any excessive computational expense.

Numerical implementation of the above
equations is straightforward in almost the
whole metallic regime. However, the metallic regime in proximity to the 
Mott transition
and the Mott insulating
regime is trickier. We will illustrate this point and its
resolution for the Mott insulator here and for the correlated metal
in the appendix. 
The retarded host Green's function 
may be separated into a singular and
a regular part ($\mc G^{\mr {reg}}$).
\be \mc G\omb=\sum_i\f{\al_i}{\om-\om_i+ i\eta} + \mcG^{\mr{reg}}\omb, 
\quad\eta\ra
0+, \quad\al>0 \ee
The sum in the above equation is over the poles $\om_i$ or the singularities
of the $\mc G\omb$ and $\al_i$'s are the corresponding
weights or residues. 
It is easily seen that if $\mc G\omb$ has
to satisfy the self-consistent equations of
DMFT [\eref{eq:local_gf}-\eref{eq:dyson},\eref{eq:Dsig}-\eref{eq:kkt}], the
residue $\al_i$ of the singularities must
satisfy self-consistent equations. 
For example, in the Mott insulator case where 
there is just one pole with residue $\al$ at the Fermi level (this
follows from the singular behavior of the self-energy and 
host Green's function in the atomic 
limit ($\ts=0$)~\cite{hubbard:1,gebhard,georges:rvw}), 
we get a cubic equation (at $T=0$), given by
\be\label{eq:alpha} {\al}^{-1}=
1+\f{4M_2}{U\sq\al\cu} \ee
where $M_2$ is the second moment of
$D_0(\om)$ [$M_2=\nint d\om\, \om\sq D_0\omb$]. 
Solutions of this equation
by Cardano's method~\cite{cardano} shows that out of three
roots, the physically reasonable root (such
that $\al\ra 1$ as $U\ra\infty$) exists
only for $U>U_{c1}\simeq 3.67t_*$ for the HCL.
For a general $U$, we solve the above cubic equation
numerically to get $\alpha$.
In the strongly correlated metallic regime,
poles can arise in the host Green's function $\mcG\omb$ even though the 
spectral density of the interacting Green's function shows a FL behavior
 within a width of low energy scale (described as
$\om_L$ later). Two poles occur symmetrically about the Fermi level 
at $\pm \om_0$ proportional to the square-root quasiparticle
weight $Z$ (see Eq.~\eref{eq:polevanish}). The Mott transition from the metallic regime 
to the insulating regime occurs through the collapse
of these two poles at the Fermi level into one single pole characteristic
of the insulating regime (as described above).
The poles of the $\mcG\omb$ lead to divergences at $\pm 3\om_0$ in the 
$\Sig\omb$ (through Eq.~\eref{eq:Dsig}). Using such a pole structure of the
Green's functions and the self-energy, the critical $U$ at which the metal
transforms into the insulator, i.e $U_{c2}$ may be estimated as 
$4.77\ts$ (see Appendix~\ref{sec:appendix}) for the HCL. We also found the same for the 
Bethe lattice ($U_{c2}=3.28\ts$) which agrees with the numerical
estimation in Ref.~\citation{ipt:2}.

At finite temperatures, the singularity at
the Fermi level must have a finite width.
However, this width could be
exponentially small in practice, 
and it is next to
impossible to capture such sharp resonance
numerically. So we utilize the spectral weight sum rule
to compute the weight of the singularity, i.e. $\alpha$.
The presence of this singularity is numerically detected
by a significant deviation (in practice 2\%) of the 
integrated spectral weight of $D_{\cal{G}}$ from unity.
The sharp resonance at the Fermi level in $D_{\cal{G}}$ is then
numerically cut off to get $\mc{D}^{\mr{reg}}_{\cal{G}}$
and the weight $\al$ is obtained by
$\al= 1-\nint d\om\,\mc{D}^{\mr{reg}}_{\mcG}$.
With the above separation of $D_{\cal{G}}$ into regular and singular
parts, the self-energy expression reduces to  
\be \Sig\omb=\Sig^{\rm reg}(\omega) + \f{U\sq\al\cu}{4\omp} \ee
where $\Sig^{\rm reg}$ is obtained through the Kramers-Kronig transform of
$D_\Sig^{\rm reg}$ which is given by
\blgn
D_\Sig^{\rm reg}\omb &=U\sq\bl
\nint d\om'\,D^{\rm{reg}}_\mcG(-\om')\nf(\om')\chi^{\rm{reg}}(\om+\om')
+\f{\al\sq}{4}\nf(-\om)D^{\rm{reg}}_\mcG(\om)
+\f{\al}{2}\chi^{\rm{reg}}(\om)\non\\
&+
\nint d\om'\,
D^{\rm{reg}}_\mcG(-\om')\nf(-\om')\chi^{\rm{reg}}(-\om-\om')
+\f{\al\sq}{4}\nf(\om)D^{\rm{reg}}_\mcG(\om)
+\f{\al}{2}\chi^{\rm{reg}}(-\om)\br
\elgn
and
\be
\chi^{\rm{reg}}(\om)=\nint d\om'\, 
D^{\rm{reg}}_\mcG(\om')D^{\rm{reg}}_\mcG(\om+\om') \nf(\om')\nf(-\om-\om')
+\al D^{\rm{reg}}_\mcG(\om)\nf(-\om)
\ee

Now we proceed to discuss our results.

%------------------------------------------------------------
%------------------------------------------------------------
\section{Results and discussions}

We have implemented the above mentioned formalism and have computed
real frequency spectral functions, dc and optical conductivities.
    
%-------------------------------------------------------------
\subsection{Spectral and transport properties}

\begin{figure}[h]
\includegraphics[height=4cm,clip]{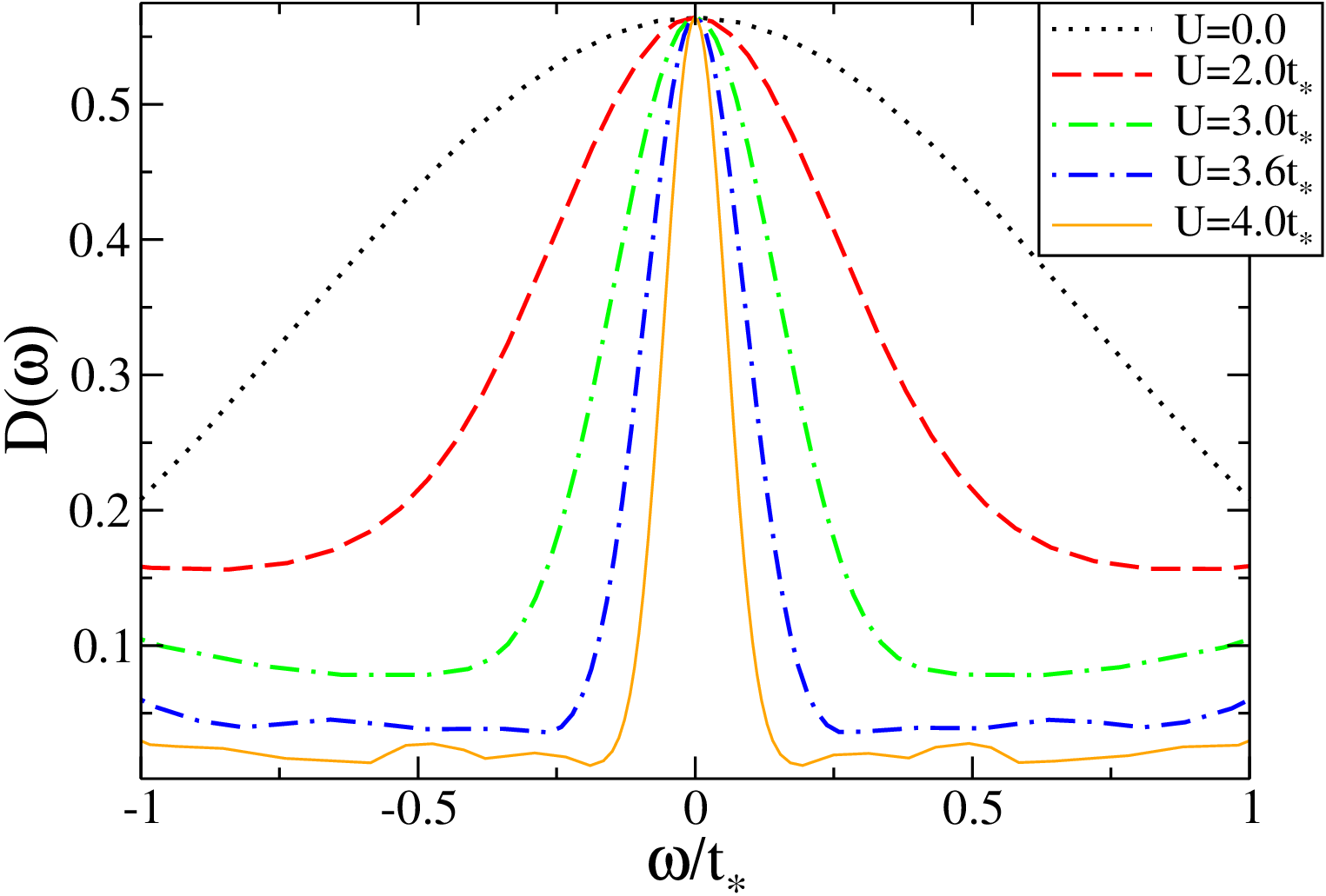}
\includegraphics[height=4cm,clip]{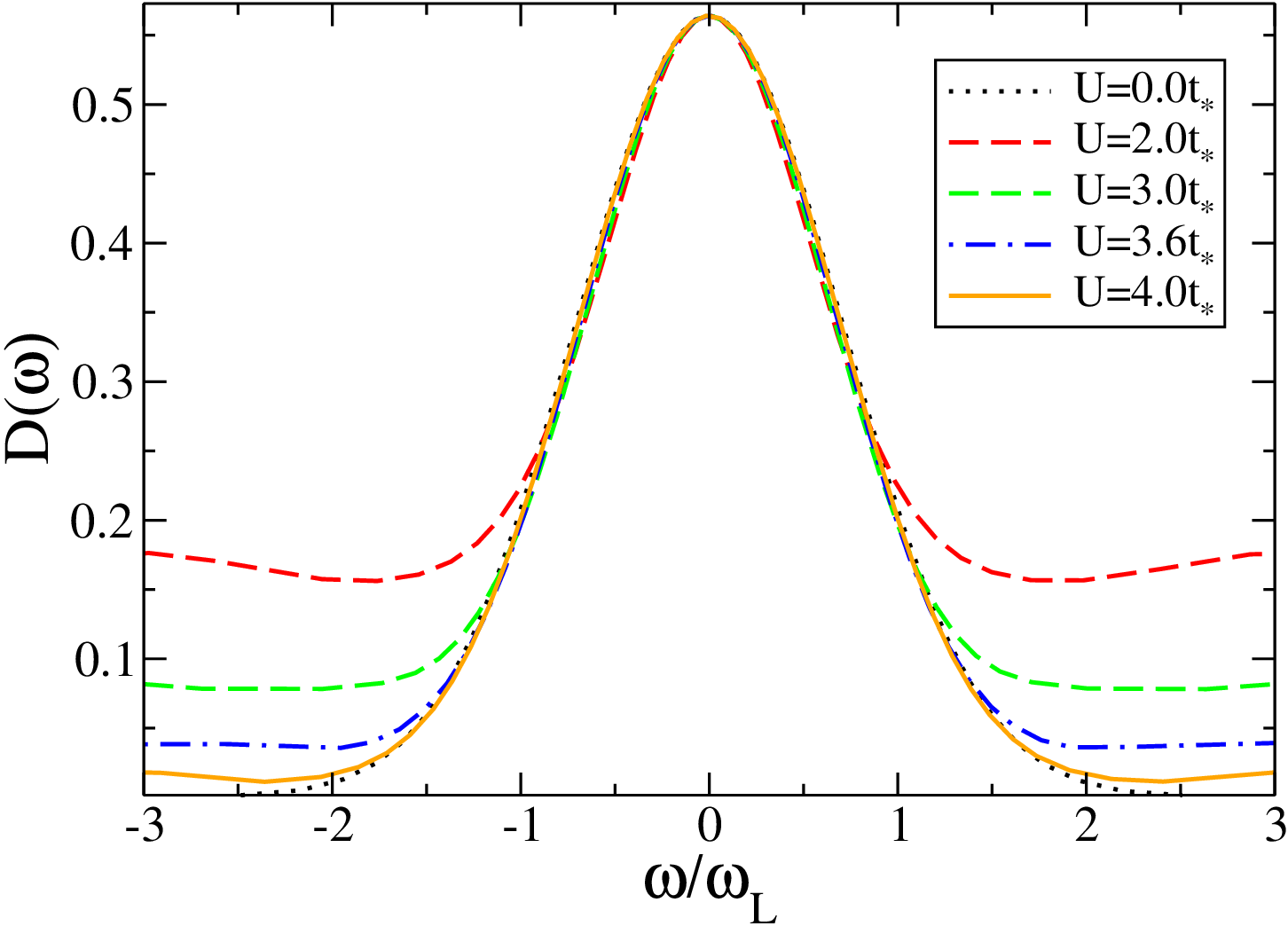}
\caption{(Color online) Spectral functions for various $U$ showing quasiparticle 
resonances at Fermi level (left). A simple $x$-axis re-scaling by $Zt_*$
leads to a collapse of all of these onto the non-interacting spectral function
in the neighbourhood of the Fermi level thus signifying adiabatic 
continuity and FL behavior.
} 
\label{fig:spf_scl}
\end{figure}
\begin{figure}[h]
\centering\includegraphics[scale=0.4,clip]{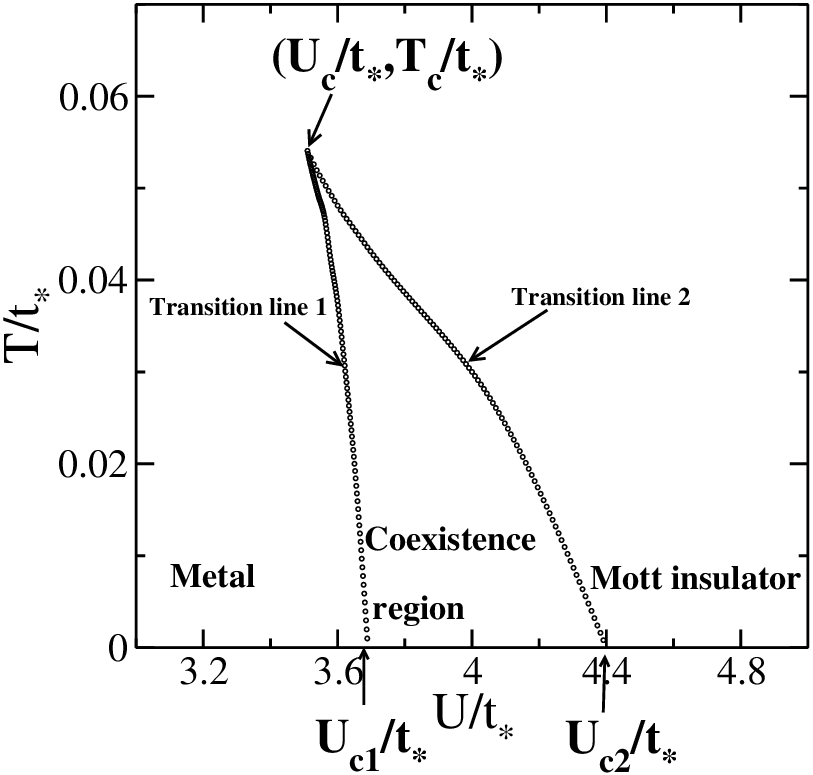}
\caption{The $U-T$ `phase-diagram' as calculated within IPT is shown
here. The solid lines mark the spinodals of the coexistence region where
the metallic and the insulating solutions coexist. 
} 
\label{fig:phase}
\end{figure}

Since our implementation is new, we would like to benchmark it by comparing
our results with other implementations. We begin with the local density 
of states (DoS) as given by $D(\om)=-{\rm Im}G(\om)/\pi$. In \fref{fig:spf_scl} we show the $D(\om)$ computed at temperature $T=0$ for various $U$ in the metallic regime.
The left panel shows the spectral function on `non-universal' scales, i.e.
{\it vs.} $\omega/t_*$, where the Hubbard bands are seen to form with increasing
$U $, while the Abrikosov-Suhl resonance at the Fermi level gets narrower.
The zero frequency value of the spectral function is also seen to be pinned
at the non-interacting ($U=0$) value. Such behavior finds a natural explanation in 
Fermi liquid (FL) behavior. A simple linear expansion of the self-energy
as 
$
\Sigma(\omega)=\Sigma(0) + \left(1-\frac{1}{Z}\right)\omega + {\cal{O}}
(\omega^2)
$
when used in Eq.~\eref{eq:gf_ht} gives $G(\om)=G_{U=0}(\om/\om_L)$ 
where $\om_L=Zt_*$
is the low energy Fermi liquid scale. Since the quasiparticle weight $Z$
decreases with increasing $U$, thus signifying an increase in effective mass
($m_*=m/Z$), the fequency width at half maximum (FWHM) for a hypercubic lattice that is given by
$\Delta=2\om_L\sqrt{\ln(2)}$ would decrease with increasing $U$. Another
inference from such low frequency scaling behavior is the universality
of the low frequency part of the spectral function. On the right panel
of \fref{fig:spf_scl}, the same spectra as the left panel are plotted
as a function of $\om/\om_L$, and they are all seen to collapse onto the
non-interacting limit spectra in the neighbourhood of the Fermi level.
All of the above behavior is of course well known and well understood.
We nevertheless emphasize that the extent of the FL regime is very small
in strong coupling, because as the right panel shows, the scaling collapse
is valid for $\om\lesssim\om_L$, where $\om_L$ is expected to decrease
exponentially with increasing $U$. The approximation of IPT does not capture
such exponential decrease of $\om_L$, instead predicting an algebraic
decrease. Nevertheless, the qualitative behavior from IPT indeed corroborates 
with other methods such as QMC.
IPT has been shown to capture the first order Mott metal-insulator 
transition qualitatively and in reasonable agreement with other techniques such
as NRG~\cite{nrg}. In the $T-U$ plane, the metallic and insulating solutions are
known to coexist in a certain region bounded by spinodal lines. The $T=0$
bounds are denoted by $U_{c1}$ and $U_{c2}$. This first order
coexistence region may be simply found by computing the 
temperature or $U$ dependence of the Fermi level
density of states ($D(0)=-{\rm Im}G(0)/\pi$) for various values of
$U/t_*$ or $T/t_*$.
The resulting phase diagram is shown below in 
\fref{fig:phase}. It agrees well 
with  those reported previously~\cite{georges:rvw,ipt:2}.

\begin{figure}
\centering
\includegraphics[scale=0.4,clip]{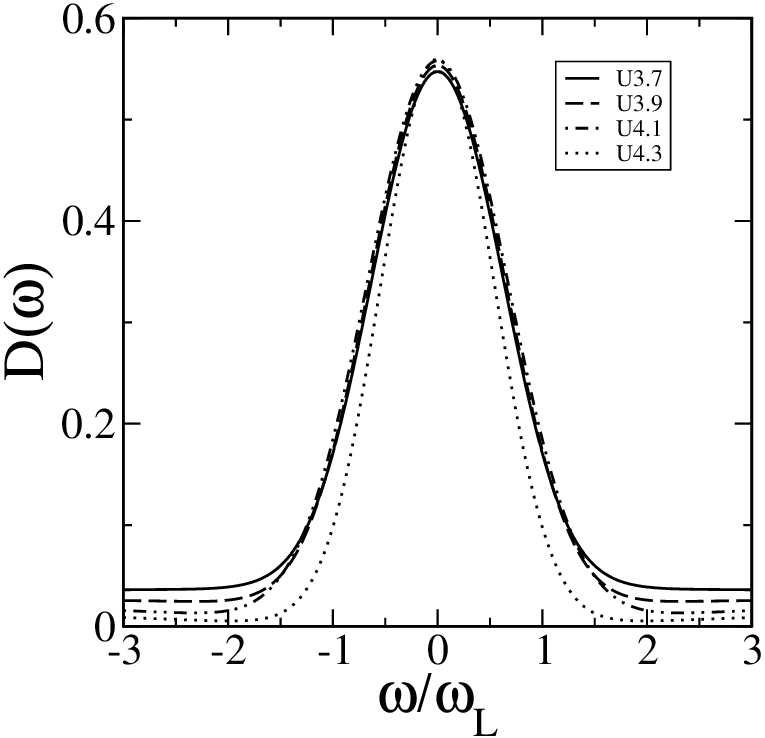}
\caption{Temperature scaling violated.
as a function of temperature for $U/\ts=2.6$ (dotted), 2.8 (dashed), 3.0 
(dotted and dashed), 3.2 (doubly dotted and singly dashed),
 3.4 (full line). }
\label{fig:tdos}
\end{figure}

The metallic region is a strongly renormalized FL, and thus
the properties close to the Fermi level must be governed by a single
low energy scale $\om_L$.
For universality to hold in the strong coupling region, the spectral function
must have the following form:
\begin{equation}
D(\omega;T)=f\left(\frac{T}{\om_L};\frac{\om}{\om_L}\right)
\end{equation}
i.e. it must be a function purely of $\tilde{T}=T/\om_L$ and 
$\tilde{\om}=\om/\om_L$~\cite{hewson}.
In \fref{fig:tdos}, we show the spectra in strong coupling for
fixed $\tilde{T}=0.2$ and increasing $U/t_*$. We see that a scaling collapse 
does not occur implying that the above universal form does not describe
the finite temperature IPT results. This non-universal behavior is an artefact of the specific iterated perturbation theory ansatz for the self-energy which is known to yield a non-universal form for its 
imaginary part~\cite{georges:mott}, 
namely ${\rm Im}\Sigma_{\rm IPT}(\om=0)\propto U^2T^2/t_*^3$.

\begin{figure}[h]
\centering
\includegraphics[scale=0.4,clip]{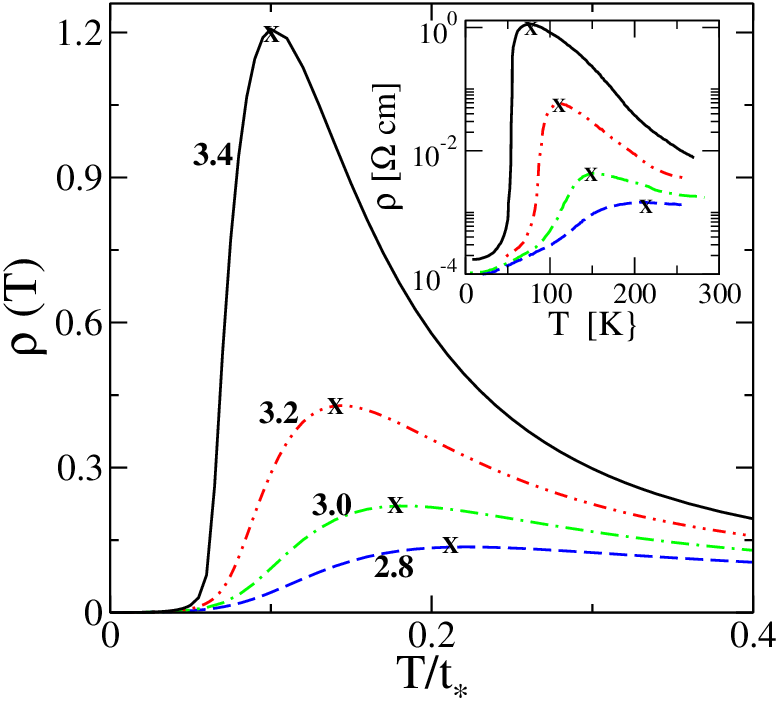}
\includegraphics[scale=0.4,clip]{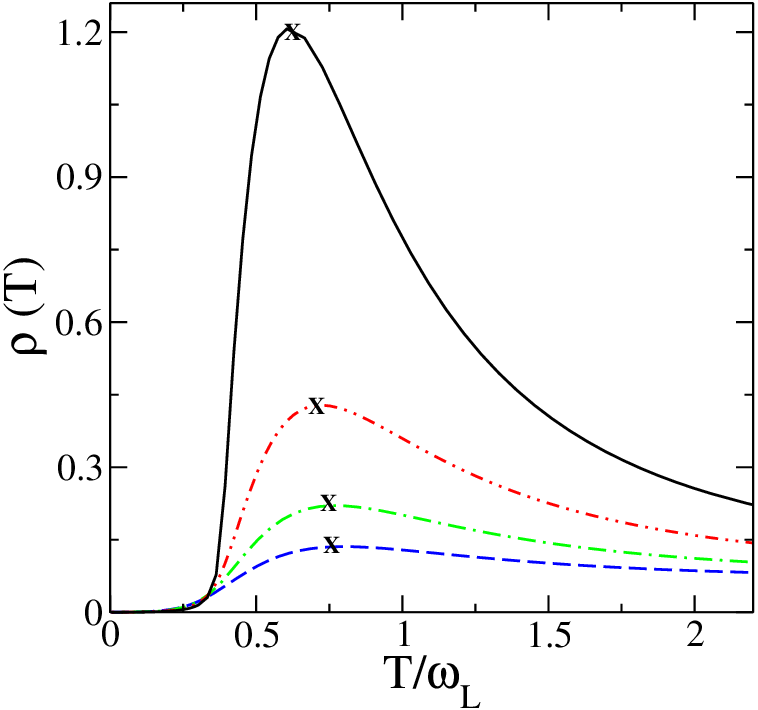}
\caption{(Color online) Left panel: Theoretically computed resistivity 
as a function of temperature for various $U/\ts$ values (indicated as
 numbers). The crosses indicate the position of the coherence
peak. Left panel inset: Experimentally measured~\cite{nis2data:1,nis2data:2} resistivity for 
NiS$_{2-x}$Se$_x$  as a function of 
pressure (indicated as numbers).
Right panel: The same resistivities as in the left panel with the
temperature rescaled by the low energy coherence scale $\om_L=Zt_*$
showing that the coherence peak is indeed a universal feature of the 
strongly correlated metallic regime.  }
\label{Fig:res_fl1}
\end{figure}

The dc resistivity is computed using \eref{eq:dccond}. Again, the
pathologies in the imaginary part of the self-energy are reflected
in the low temperature FL region in the following way. 
Although the resistivity
does have a $T^2$ form, the expected universal scaling form of $(T/\om_L)^2$
is not obtained~\cite{georges:mott}. However, surprisingly, the 
`coherence peak' position, 
which represents a crossover between low temperature coherent behavior
to high temperature incoherent behavior does
seem to be a universal feature in strong coupling as seen in 
\fref{Fig:res_fl1} occurring at $\tilde{T}=0.6$. The crosses represent the peak
position, and as is seen in the right panel, these crosses line up at
a single $\tilde{T}$. Thus the position
of the coherence peak may be used to infer the low energy scale
in a real material.
In the inset of the left panel, we show the experimentally 
measured~\cite{nis2data:1,nis2data:2} resistivity of NiS$_{2-x}$Se$_x$  as a function of
pressure in Kbar (indicated as numbers). The resistivity for the lowest pressure
rises dramatically with increasing $T$, before reaching a coherence maximum,
 and then decreasing slowly for higher temperatures. With increasing 
pressures, the initial rise becomes more gradual, and the coherence peak
shifts to higher temperatures. An increase in pressure leads to a decrease
in lattice spacing, thus increasing $t_*$, the hopping parameter, while the
local Coulomb repulsion $U$ remains unaffected. Thus
increasing pressure can be interpreted as a decrease in the $U/t_*$ ratio.
A comparison of the inset with the main figure of the left panel
clearly indicates qualitative agreement. The initial rise of $\rho(T)$
with $T$ is much sharper in experiment than in theory, but the
rest of the features, including a shift of the coherence peak to higher
temperatures with increasing pressure, are indeed observed. We emphasize
here that the agreement is only qualitative and as such, no attempt is made 
to obtain quantitative agreement.

We now study the resistivity hysteresis as obtained within IPT
approximation.

%-----------------------------------------------------------------
\subsection{Thermal hysteresis}

\begin{figure}[h]
\centering
\includegraphics[scale=0.20,clip]{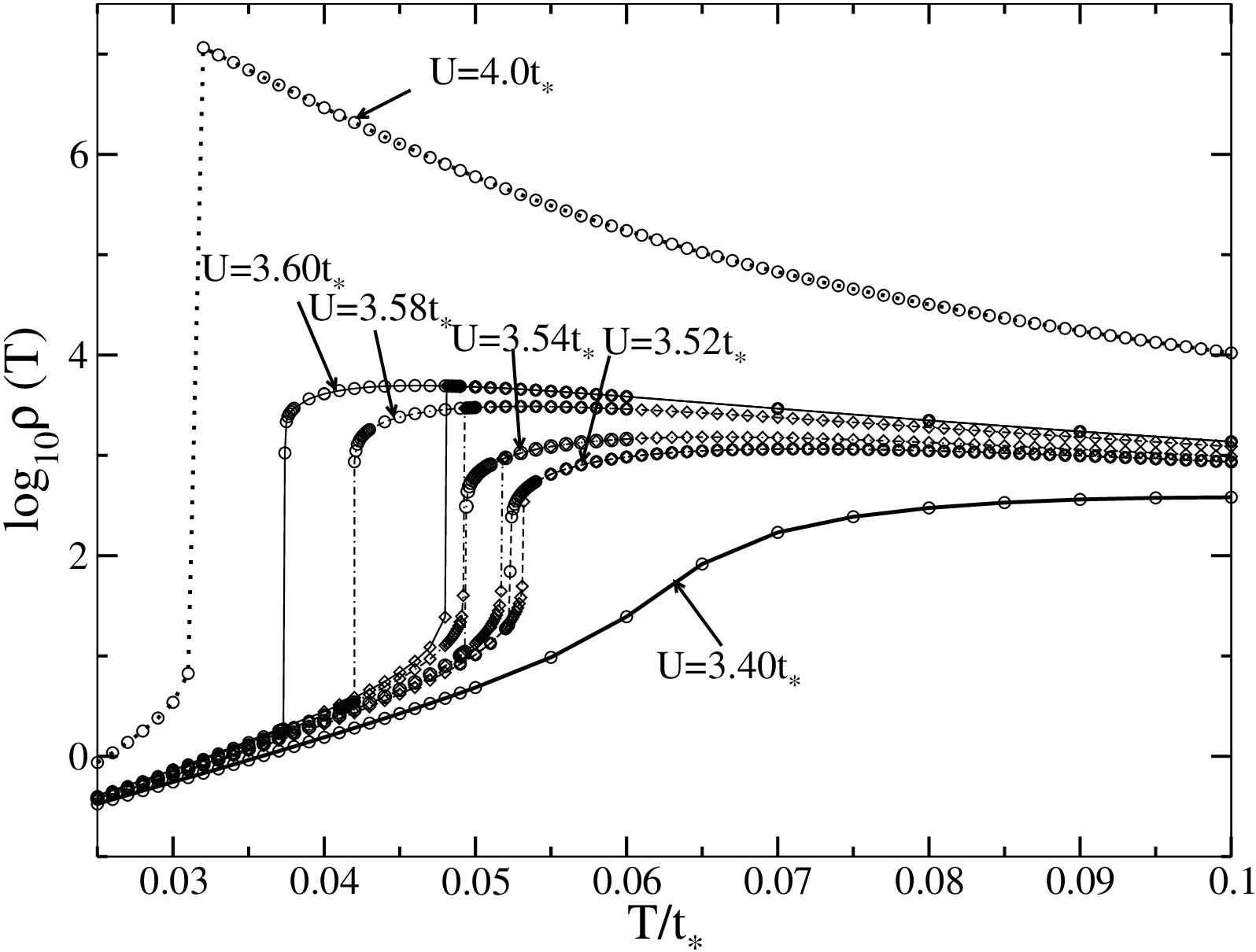}
\includegraphics[scale=0.20,clip]{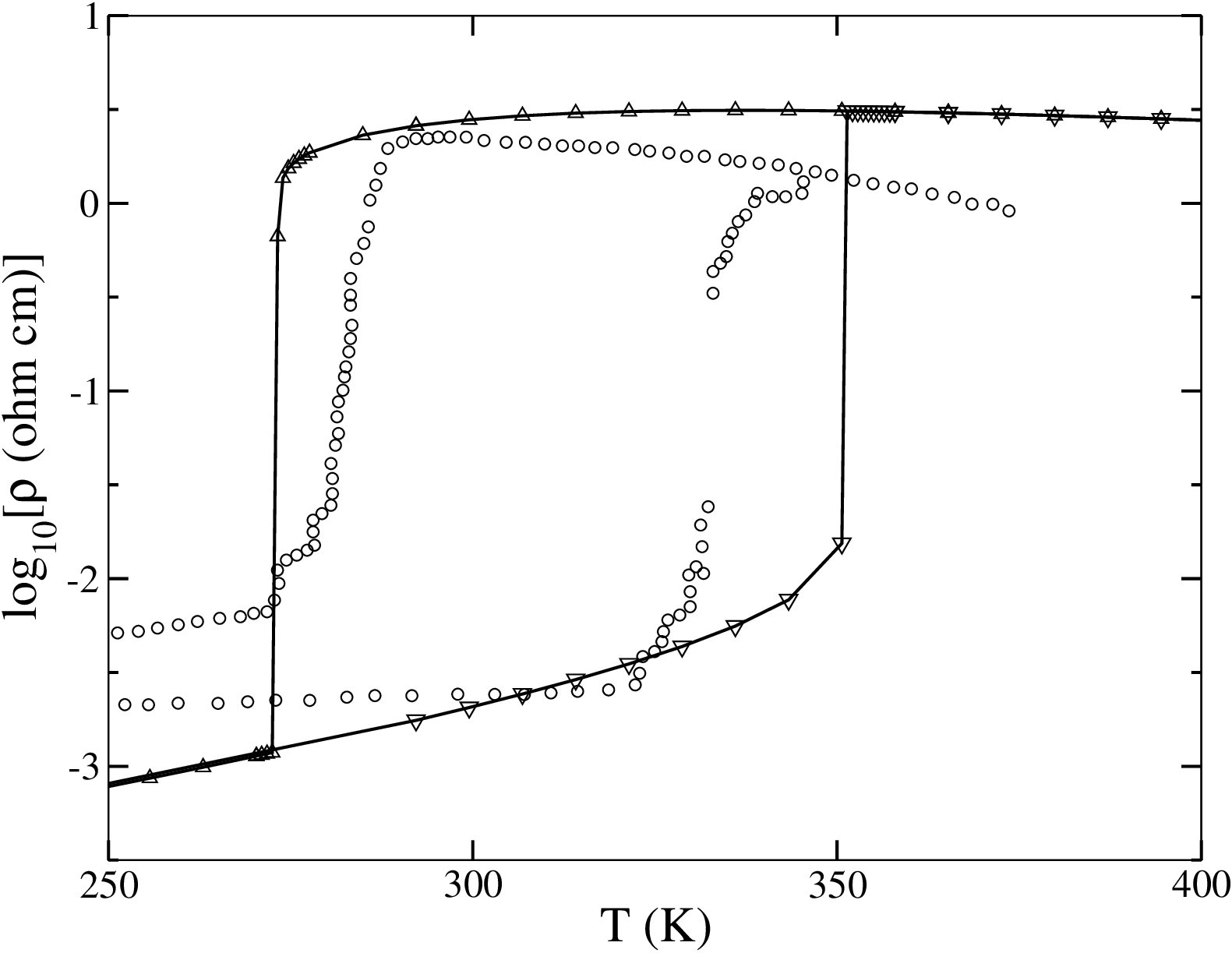}
\caption{Left panel: The resistivity on a log-scale exhibits thermal hysteresis
as a function of temperature when  $U_c<U<U_{c1}$ (see text for discussion). 
For $U=3.4t_*$ second order transition is observed without hysteresis.
For $U=4.0t_*$ hysteresis is not observed though the metal-insulator transition
is of first order. 
\noindent
Right panel: Comparison of theoretically obtained thermal hysteresis ($U=3.6t_*$)
with experimental observations in V$_2$O$_3$~\cite{kuwamoto}.}
\label{fig:hyst}
\end{figure}

\Fref{fig:hyst}  shows the thermal hysteresis in resistivity obtained 
through heating and cooling cycles for fixed $U$'s ($\ts=1$) in the coexistence region.
The area enclosed by the hysteresis loop decreases as $U\rightarrow U_c$.
A full cycle hysteresis is observed only in the 
region $U_c (\sim 3.5) < U < U_{c1} (\sim 3.7)$. 
For $U< U_c$, the metal-insulator transition is continuous
and hence of second order while for  $U> U_{c1}$ the transition is 
discontinuous but
hysteresis is not obtained. 
The hysteresis can be qualitatively explained through the coexistence region in \fref{fig:phase}.
For $U_c<U<U_{c1}$ the $T=0$ ground state is a FL. As one increases
$T$ for a fixed $U$, and crosses the spinodal on the right (transition line 2), a first order
transition to a paramagnetic insulating state occurs, which upon cooling
does not transit to the paramagnetic insulating state until the left spinodal
is crossed (transition line 1). Thus as one crosses the coexistence region and reenters via
heating/cooling, thermal hysteresis would be obtained. 
The hysteretic behavior seen in the present theory is naturally very 
far removed from the rich experimentally observed hysteresis seen e.g. in
V$_2$O$_3$. The coexistence of metallic and insulating islands
has been experimentally observed in thin films~\cite{grygiel}, as well
as in manganites~\cite{ddsarma},
while within DMFT, where spatial 
inhomogeneities are completely ignored, the coexistence
is just that of the metallic and insulating solutions. Experimentally, the 
resistivity does not increase monotonically with heating or cooling. Instead,
 multiple steps or 
avalanches are observed to accompany the hysteretic behavior. 
While such details are absent in the present theory, nevertheless, 
we carry out a direct comparison of  our hysteresis result with the 
one found experimentally in doped
V$_2$O$_3$\cite{kuwamoto} (right panel of \fref{fig:hyst}). The purpose of such a comparison being to assess
if the half-filled Hubbard model can capture at least the qualitative aspects
of real materials. If it does, then the hope would be that a realistic theory
based on finite dimensions including spatial inhomogeneities would be 
able to capture the experimentally observed behavior quantitatively. 
The best fit of the hysteresis result for a specific interaction 
($U=3.6t_\ast$), yields 
 $t_*\sim 7305\mr{K}$ $(0.63 \mr{eV})$. This agrees well
with independent bandstructure calculations 
for V$_2$O$_3$~\cite{mattheiss} and is of the right magnitude.

We now focus on optical transport in the next subsection.
%
%-------------------------------------------------------------------------------
\subsection{Optical conductivity:}

\begin{figure}[h]
\centering
\includegraphics[scale=0.5,clip]{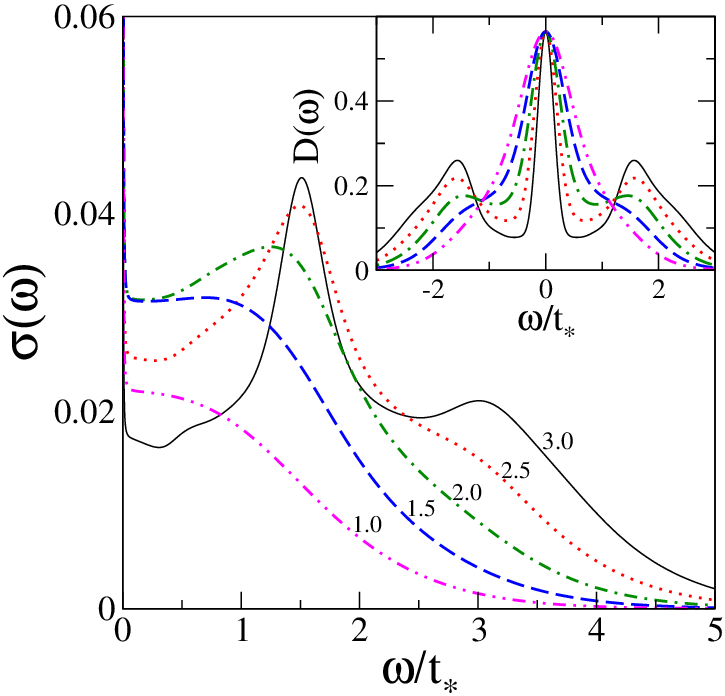}
\caption{(Color online) The main panel shows the zero temperature optical conductivity
for $U/t_*$ values ranging from 1.0 to 3.0 (indicated by numbers). The inset
shows the corresponding spectral functions.
}
\label{fig:OC_T0_U1_U3}
\end{figure}

As can be naturally expected, changes in interaction strength $U/t_*$ and
especially the Mott MIT affect dynamical or optical
 transport properties strongly. \Fref{fig:OC_T0_U1_U3} 
shows the computed $T=0$ optical conductivity $\sigma(\om)$ as 
$U/t_*$ increases from a low value of 1.0 to a moderately strong value
of 3.0. The inset shows the corresponding spectral functions. Several
very interesting features can be seen. 
As $U$ increases, a strong absorption
feature emerges at $\sim U/2$ for $U\gtrsim 2t_*$, while a second 
peak at $\sim U$ emerges beyond 3$t_*$. For low values, none of these 
features may
be distinguished. The first peak arises because of excitations between either
of the Hubbard bands and the Fermi level, while the second peak represents
excitations between the lower and upper Hubbard band. The zero frequency
Drude peak is also present, but is not visible, since it has a Dirac-delta
function form. The changes in optical conductivity are naturally understood
through the inset of \fref{fig:OC_T0_U1_U3}. For low values of $U/t_*$,
a single featureless spectral function is obtained, while the emergence
of distinct Hubbard bands in the spectra mark the emergence of the first
and second absorption peaks in the optical conductivity. At around the half bandwidth ($\om\simeq 1.15t_*$) a universal crossing point is is seen in the spectra.

\begin{figure}[h]
\centering
\includegraphics[scale=0.5,clip]{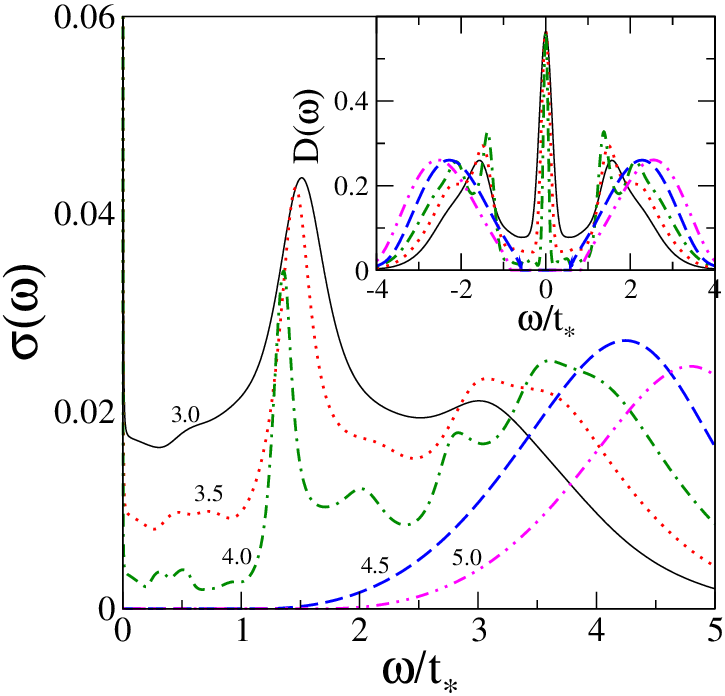}
\caption{(Color online) The main panel shows the zero temperature optical conductivity
for $U/t_*$ values ranging from 3.0 to 5.0 (indicated by numbers). The inset
shows the corresponding spectral functions.
}
\label{fig:OC_T0_U3_U5}
\end{figure}
The effect of Mott transition on the optical absorption is illustrated
clearly in \fref{fig:OC_T0_U3_U5}, which is similar to the previous
figure, except that the values of $U/t_*$ considered here increase from
3.0 to 5.0. The transition from the correlated metallic phase to the 
Mott insulator phase occurs at $U/t_*\simeq 4.5$, where, within IPT, a 
large gap $\sim U-2t_*$ is known to form. The main panel shows optical
conductivity as a function of $\omega/t_*$. In the metallic phase
($U<U_{c2}$), the first absorption peak is seen to get narrower
and surprisingly gets red-shifted as the Mott transition is approached.
The second absorption peak begins to dominate as $U\rightarrow U_{c2}$
and becomes the sole feature in the Mott insulating phase. The $U>U_{c2}$
optical conductivity is seen to possess a clear optical gap, which increases
with increasing $U$ and reflects
the presence of the gap in the density of states (see inset).

\begin{figure}[h]
\centering
\includegraphics[scale=0.35,clip]{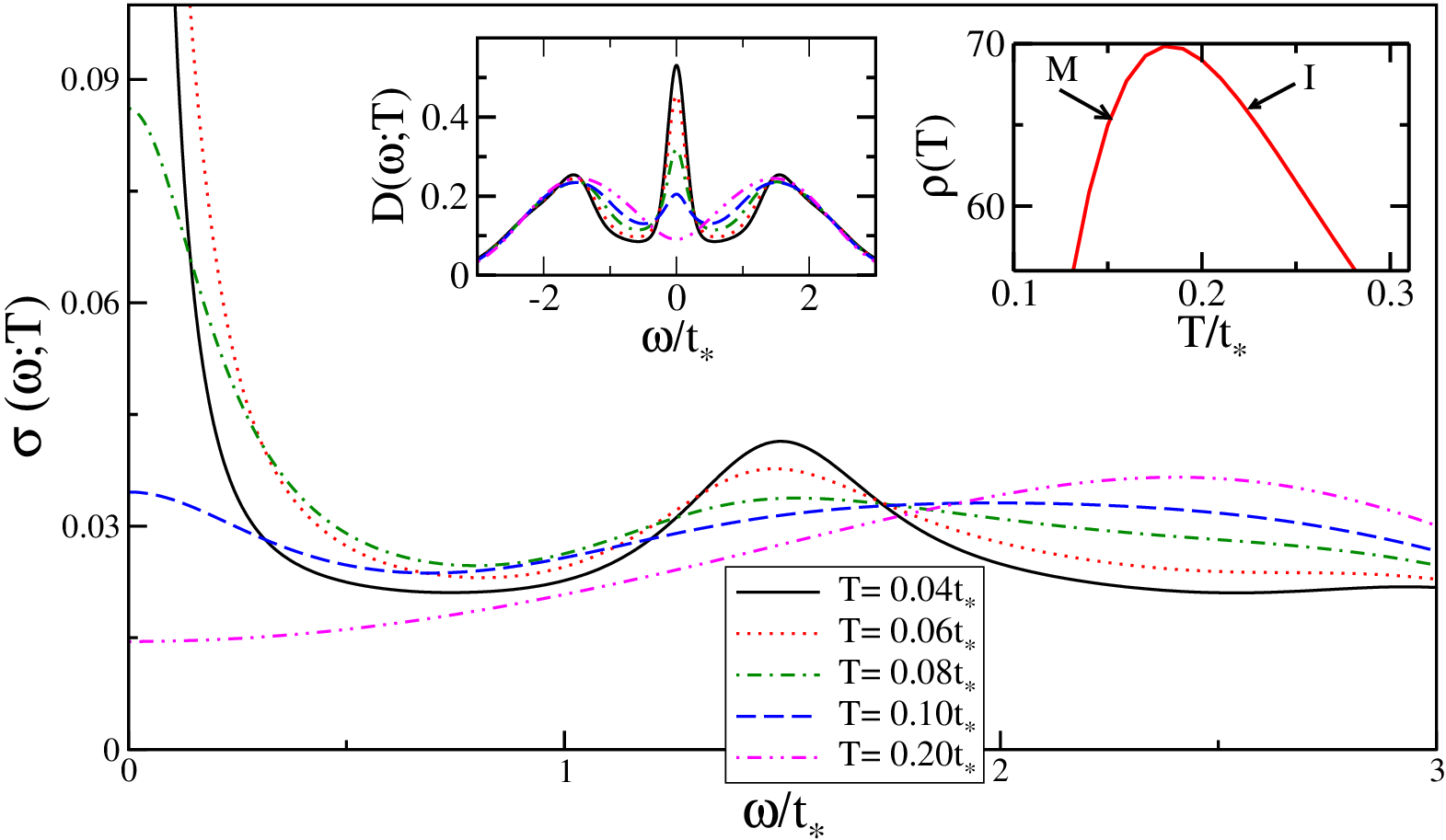}
\caption{(Color online) The main panel shows the temperature evolution of optical conductivity
for $U/t_*=3.0$. The left inset
shows the corresponding spectral functions and the right inset shows the resistivity at $T=0.2\ts$. M marks for metallic ($d\rho/dT>0$) and I marks for insulating ($d\rho/dT<0$) behavior.
}
\label{fig:OC_U3}
\end{figure}

The temperature evolution of the optical conductivities is equally
interesting. In \fref{fig:OC_U3} we show the $\sigma(\omega;T)$
behavior for various $T$'s for a moderately strong
interaction strength of $U=3.0t_*$. The inset again
shows the corresponding spectral functions. 
The first absorption peak at lower temperatures loses spectral weight 
as temperature increases, which is gained by the second peak, and again
an almost universal crossing point is seen, marking the frequency across
which the transfer of spectral weight occurs. The Drude peak at 
$\om=0$ diminishes in height, consistent with the 
dc conductivity values, and disappears completely
at $T=0.2$ by forming a shoulder-like feature at higher frequency 
$\om\sim 2.5\ts$.

An earlier investigation\cite{mutou} (though it was not exactly calculated for the half-filled HM) claimed that the shoulder formation is actually shifting    
of Drude peak (pseudo-Drude peak) and the phase is metallic. However, here 
we argue that this arises due to the pseudogap formation in the DoS 
(see the left inset)
and it is actually an insulating state since $d\rho(T)/dT < 0$ (see the right inset) and the phase lies in the crossover regime.

The  universal crossing point, known as the \emph{isosbestic point} in the 
metallic phase  has been
observed in several materials, e.g. V$_2$O$_3$~\cite{baldassare:V2O3}, 
NiS$_{2-x}$Se$_x$~\cite{perucchi:NiSe}, 
La$_{2-x}$Sr$_x$CuO$_4$~\cite{uchida:LSCO}, La$_{1-x}$Sr$_x$TiO$_3$~\cite{okimoto:LSTO}. Occurrence of isosbestic points are not well-explained. Nevertheless such a point is believed to have a close connection with $f$-sum rules and its location is associated to microscopic energy scales in correlated systems~\cite{vollhardt:2,freericks}.  
Here, we see that the dynamics also exhibit a similar feature indicating
an even more general basis for its existence. If we use our earlier 
estimation of $\ts$ (0.6 eV) for V$_2$O$_3$ or
the LDA-calculated value, we find that
the isosbestic point arises at $\sim$ 0.9-1.0 eV which lies in
the mid-infrared range, and which is close to that seen in 
a recent infra-red spectroscopy measurement~\cite{baldassare:V2O3} i.e.\ at 
6000 cm$^{-1}$ ($\sim 0.7$ eV). 
\begin{figure}[h]
\centering
\includegraphics[scale=0.5,clip]{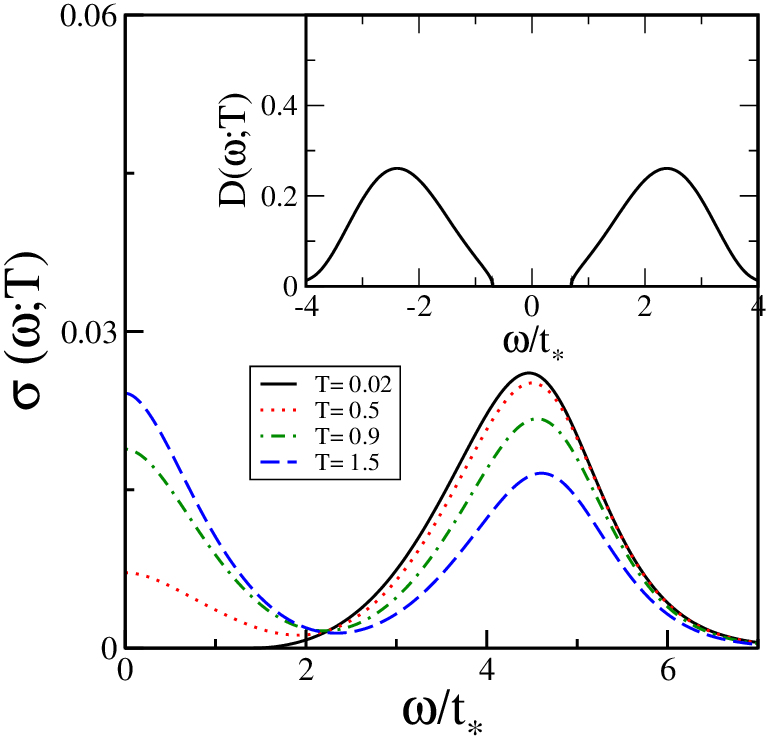}
\caption{(Color online) The main panel shows optical conductivity for temperatures
$T/t_*=0.02, 0.5, 0.9$ and $1.5$. The inset shows the corresponding spectral
function which exhibits negligible changes as temperature is increased.
}
\label{fig:opt_cond3}
\end{figure}
Finally we show the optical conductivity for the Mott insulating regime
($U/t_*=4.7$) evolving with temperature in \fref{fig:opt_cond3}.
A single absorption peak is seen at low temperatures, and as $T$ increases,
spectral weight is transferred from this peak to lower frequencies, and the
absorption peak diminishes, and experiences a blue shift. The spectral function
exhibits negligible change as a function of temperature and although, there 
is indeed, an exponentially small rise in the  density of states in the neighbourhood
of the Fermi level, it visually appears to coincide with the $T=0$ DoS.
%--------------------------------------------------------------------------
\section{Conclusion}

A systematic study of the half-filled Hubbard model within dynamical mean field
theory is carried out, with the focus being on universality, scaling
and qualitative comparison to experiments.
We reformulate a well-known and extensively employed impurity solver for
the effective impurity problem that the lattice problem gets mapped onto 
within DMFT, namely the iterated perturbation theory, such that some of
the problems with previous implementations  have been overcome.
We find that the coherence peak in the resistivity is a universal feature.
A comparison with experimental measurements of resistivity in Se doped NiS$_2$
with varying pressure yields qualitatively excellent agreement.
Thermal and pressure driven hysteresis is shown to be qualitatively
explicable within this scenario, and again, a comparison of thermal hysteresis
with experiments in V$_2$O$_3$ are seen to yield a reasonable number for the
hopping integral. The transfer of spectral weight across the Mott transition
and the isosbestic points have been highlighted in the study of optical
conductivity. We conclude that the Hubbard model does indeed represent
an appropriate phenomenological model that can qualitatively explain
a large range of phenomena observed in transition metal oxides. This offers
hope for more detailed material specific studies such as those
employing LDA+DMFT~\cite{ldadmft:1,ldadmft:2,ldadmft:3} approaches to obtain quantitative agreement with experiments.

\section*{Acknowledgement}
The authors would like to acknowledge funding and support
from the DST and JNCASR. H.B. owes to Sudeshna Sen and Nagamalleswara Rao Dasari for their help in proof-reading and to Pinaki Majumadar for useful discussions.

\appendix
\section*{Appendix A: Analytical calculation of \texorpdfstring{$U_{c1}$}{U1} and \texorpdfstring{$U_{c2}$}{U2} at \texorpdfstring{$T=0$}{T} }
\label{sec:appendix}

In this section, we demonstrate that a pole structure ansatz for the
Green's functions and self energy combined with the IPT equations
yields $U_{c1}$ and $U_{c2}$, the bounds
of the $U$ interval where the metallic and insulating solutions coexist at
zero temperature.

If $\mcG$ has simple poles at $\pm \om_0$ with equal residue $\alpha/2$ (due to symmetry), then
the singular part of  $\mcG$ may be expressed as 
\be \mcG\omb=\f{\al}{2}
 \bl \f{1}{\om-\om_0}+\f{1}{\om+\om_0}\br
=\al\f{\om}{\om\sq-\om\sq_0}
\ee

\noindent
Now using the Dyson equation (Eq.~\eref{eq:dyson}) and the moment expansion of the Hilbert transform,
we get
\bes
\Sig\omb=\f{1}{\al}\f{\om\sq-\om\sq_0}{\om}
-\om+\Sig\omb-\f{M_2}{\om-\Sig\omb}
\ees
%where $M_2$ is the second moment of the non-interacting density of states $D_0\omb$.

\noindent
Thus
\be
\Sig\omb=\om-\f{M_2}{\f{1}{\al}\f{\om\sq-\om\sq_0}{\om}-\om}
\ee
Therefore poles occur at
\be\label{eq:sigpole} \om'_0=\pm \f{\om_0}{\sqrt{1-\al}} \ee

\noindent
Using the structure of $\mcG$ in the IPT equations, it is easy to see that
poles at $\pm\omega_0$ in  $\mcG$
will give rise to poles in the self energy at $\pm\omega_0$ and 
$\pm 3\omega_0$. Only the latter are physically acceptable, since the former
leads to a self-consistent value of the residue $\alpha$ equal to $0$. So 
when $\om'_0=\pm 3\om_0$, $\al$ turns out to be 8/9.

\noindent
The singular part and the FL part of the self energy may be
expressed in a combined way as
\be
\Sig\omb=\f{U\sq \al\cu}{8}
\bl \f{1}{\om-3\om_0}+\f{1}{\om+3\om_0}\br-\left(1-\frac{1}{Z}\right)\om
\ee
%where $Z$ is the quasiparticle weight.

\noindent
This gives 
\be
\Sig(\om_0)=-\f{U\sq \al\cu}{32\om_0}-\lb 1-\frac{1}{Z}\rb \om_0
\ee
Now $\mcG\inv(\om_0)=0$.
So the Dyson equation (Eq.~\eref{eq:dyson}) gives
\be \Sig(\om_0)=-G\inv(\om_0)=-\g(\om_0)-\f{M_2}{\g(\om_0)}=-\om_0+\Sig(\om_0)-\f{M_2}{\om_0-\Sig(\om_0)}\ee
and hence, after rearranging we get
\be
\label{eq:polevanish}
\om\sq_0=Z\lb M_2-\f{U\sq\al\cu}{32}\rb
\ee
The above result shows that the pole position is proportional to the 
square root of the quasiparticle weight.
The transition from the metallic to the insulating regime occurs when 
$\om_0=0$.This occurs when $M_2=\f{U\sq\al\cu}{32}$, i.e. for HCL, with
$\al=8/9$,
\be \f{U_{c2}}{\ts}=\f{4}{\al^{3/2}}=\f{27}{4\sqrt 2}\ee

\noindent
To get $U_{c1}$, we refer back to  Eq.~\eref{eq:alpha} which holds
in the insulating phase. This self consistent equation for the pole
residue in the insulating phase $\alpha$ has real roots only when
$U\sq \ge 27 M_2$, i.e. for HCL
\be U_{c1}=\f{3\sqrt{3}}{\sqrt 2}\ts\ee
The above analysis shows that we can analytically estimate $U_{c1}$
and $U_{c2}$ within the same framework.

%\section*{References}
%\input{refs}
%\bibliographystyle{unsrt}
%\bibliographystyle{iopart-num}
\bibliographystyle{prsty}
\bibliography{ref}
\end{document}